\input harvmac
\input amssym.tex
\input psfig
\newcount\figno
\figno=0
\def\fig#1#2#3{
\par\begingroup\parindent=0pt\leftskip=1cm\rightskip=1cm\parindent=0pt
\global\advance\figno by 1
\midinsert
\epsfxsize=#3
\centerline{\epsfbox{#2}}
\vskip 12pt
{\bf Fig. \the\figno:} #1\par
\endinsert\endgroup\par
}
\def\figlabel#1{\xdef#1{\the\figno}}
\def\encadremath#1{\vbox{\hrule\hbox{\vrule\kern8pt\vbox{\kern8pt
\hbox{$\displaystyle #1$}\kern8pt}
\kern8pt\vrule}\hrule}}
\def\underarrow#1{\vbox{\ialign{##\crcr$\hfil\displaystyle
 {#1}\hfil$\crcr\noalign{\kern1pt\nointerlineskip}$\longrightarrow$\crcr}}}
%
\overfullrule=0pt

%
\def\tilde{\widetilde}
\def\bar{\overline}

\font\zfont = cmss10 

\def\bigone{\hbox{1\kern -.23em {\rm l}}}
\def\ZZ{\hbox{\zfont Z\kern-.4emZ}}

\Title{hep-th/0403199} {\vbox{\centerline{Parity Invariance }
\bigskip
\centerline{For Strings In Twistor Space}}}
\smallskip
\centerline{Edward Witten}
\smallskip
\centerline{\it Institute For Advanced Study, Princeton NJ 08540 USA}

\bigskip
\bigskip
Topological string theory with twistor space as the target makes
visible some otherwise difficult to see properties of perturbative
Yang-Mills theory.  But left-right symmetry, which is obvious in
the standard formalism, is highly unclear from this point of view.
Here we prove that tree diagrams computed from connected
$D$-instanton configurations are parity-symmetric.  The main point
in the proof also works for loop diagrams.

\noindent

\Date{March, 2004}
\newsec{Introduction}

\nref\dewitt{B. DeWitt, ``Quantum Theory Of Gravity, III:
Applications Of
The Covariant Theory,'' Phys. Rev. {\bf 162} (1967) 1239.}%
\nref\pt{S. Parke and T. Taylor, ``An Amplitude For $N$ Gluon
Scattering,''
Phys. Rev. Lett. {\bf 56} (1986) 2459.}%
\nref\bg{F. A. Behrends and W. T. Giele, ``Recursive Calculations
For Processes
With $N$ Gluons,'' Nucl. Phys. {\bf B306} (1988) 759.}%
\nref\bkd{Z. Bern, L. Dixon, and D. Kosower, ``Progress In
One-Loop QCD
Calculations,'' hep-ph/9602280, Ann. Rev. Nucl. Part. Sci. {\bf 36} (1996) 109.}%
\nref\newbern{C. Anastasiou, Z. Bern, L. Dixon, and D. Kosower,
``Planar Amplitudes In Maximally Supersymmetric Yang-Mills
Theory,'' hep-th/0309040; Z. Bern, A. De Freitas, and L. Dixon,
``Two Loop Helicity Amplitudes For Quark Gluon Scattering In QCD
and Gluino Gluon Scattering In Supersymmetric Yang-Mills
Theory,'' JHEP 0306:028 (2003), hep-ph/0304168.}%
Perturbative Yang-Mills scattering amplitudes have many unexpected
simplifications that have been found in very early studies
\dewitt, in more contemporary investigations of multi-gluon tree
level scattering \refs{\pt,\bg}, and in studies of loop diagrams
\refs{\bkd,\newbern}.

A certain topological string theory \ref\witten{E. Witten,
``Perturbative Gauge Theory As A String Theory In Twistor Space,''
} in which the target space is twistor space \ref\penrose{R.
Penrose, ``Twistor Algebra,'' J. Math. Phys. {\bf 8}  (1967) 345.}
gives a new approach to understanding some of these questions.
(For an open string version of twistor string theory, see
\ref\berko{N. Berkovits, ``An Alternative String Theory In Twistor
Space For ${\cal N}=4$ Super Yang-Mills,'' hep-th/0402045.}. For
an alternative proposal involving mirror symmetry, see
\ref\vafa{A. Neitzke and C. Vafa, ``$N=2$ Strings And The
Twistorial Calabi-Yau,'' hep-th/0402128.}. See also further
developments in \ref\newvafa{N. Nekrasov, H. Ooguri, and C. Vafa,
``$S$-Duality And Topological Strings,'' hep-th/0403167.}.)
However, while making some aspects of perturbative gauge theory
more transparent, the twistor formalism obscures other properties
such as the parity invariance or left-right symmetry of the model.

For example, tree amplitudes with all but two gluons having
positive helicity are called maximal helicity violating or MHV
amplitudes.  They are described by a simple holomorphic function
\refs{\pt,\bg} that can be readily computed in twistor space
\witten; the computation involves current algebra correlation
functions, something which is natural in view of observations made
some time ago \ref\nair{V. P. Nair, ``A Current Algebra For Some
Gauge Theory Amplitudes,'' Phys. Lett. {\bf B214} (1988) 215.}.
Parity symmetry converts these amplitudes to amplitudes with all
but two gluons having negative helicity; we call these the dual,
opposite helicity, or googly MHV amplitudes. In the standard
formalism, it is obvious that the parity conjugate amplitude is
obtained by exchanging the two types of spinors (or, for real
momenta, by simply complex-conjugating the amplitude). In the
twistor formalism, this is far from apparent, but has been shown
to be true at tree level if only connected $D$-instantons are
considered \ref\rsv{R. Roiban, M. Spradlin, and A. Volovich, ``A
Googly Amplitude From The $B$ Model In Twistor Space,''
hep-th/0402016; R. Roiban and A. Volovich, ``All Googly Amplitudes
From The $B$ Model In Twistor Space,'' hep-th/0402121.}, and also
in another approach
\ref\cjz{Chuan-Jie Zhu, ``The Googly Amplitudes In Gauge Theory,''
hep-th/0403115.} based on tree diagrams
with MHV amplitudes for vertices
\ref\csw{F. Cachazo, P. Svrcek, and E. Witten, ``MHV Vertices And
Tree Amplitudes In Gauge Theory,'' hep-th/0403047.}.

The purpose of the present paper is to make this result more
transparent and to generalize it. We will consider amplitudes with
$p$ positive helicity gluons and $q$ negative helicity gluons for
arbitrary $p,q$.  For any number of loops, we will show, in
section 2, that the twistor amplitude can be expressed, after
integrating out some of the variables and introducing some new
ones, as an integral in which the integration region is
symmetrical in $p$ and $q$.

For tree diagrams, we go farther and prove in section 3 that the
measure, as well as the integration region, is symmetric in $p$
and $q$, and therefore that the amplitudes computed in twistor
theory are parity-symmetric.   We cannot extend this analysis to
loop diagrams, because the proper definition of the integration
measure for loop diagrams in twistor space (that is, for
$D$-instanton configurations of positive genus) remains unclear.

The proof of parity invariance includes as a special case a new
method of computing the tree level dual MHV amplitudes, that is
the amplitudes with $(p,q)=(2,n-2)$, since the ``ordinary'' MHV
amplitudes with $(p,q)=(n-2,2)$ are readily computed from twistor
space.

Another argument for parity symmetry in twistor string theory has
been given in \ref\berkmotl{N. Berkovits and L. Motl, ``Cubic
Twistorial String Field Theor,'' hep-th/0403187.},
based on a Fourier transform that presumably is a real analog of
the Serre duality that we use in section 2.

As in \rsv, we consider only {\it connected} $D$-instantons.  This
poses something of a puzzle, because there is also evidence \csw\ that tree
amplitudes can be computed from completely disconnected
$D$-instanton configurations.  It is not yet clear why there is
apparently more than one way to compute Yang-Mills tree amplitudes
from curves in twistor space.

\newsec{Parity Invariance And Serre Duality}

\subsec{Wavefunctions}

We consider in $U(N)$ gauge theory the scattering of external
gluons that are described by spinors $\lambda_{i}$,
$\tilde\lambda_i$ and momenta $p_{i\,a\dot
a}=\lambda_{i\,a}\tilde\lambda_{i\,\dot a}$. A momentum eigenstate
with momentum $p_i$ is described in twistor space by a
wavefunction that is roughly
\eqn\murgo{\Psi_i(\lambda,\mu)=\bar\delta(\langle\lambda,
\lambda_i\rangle)\exp(i[\mu,
\tilde\lambda_i]).}
 The idea is that the delta function
 describes a state with $\lambda=\lambda_i$, and the plane wave
 dependence on $\mu$ describes a state with
 $\tilde\lambda=\tilde\lambda_i$.
Here essentially as in \csw,\foot{We will include a factor of
$-i$ in the definition of $\bar\delta(f)$ relative to \csw\ to
avoid having an unnatural factor in eqn. 2.2. What we here call
$\delta^2(f)$ was called $\delta(f)$ in \csw.} for any holomorphic
function $f$, the symbol $\bar\delta(f)$ represents the closed
$(0,1)$-form $-id\bar f\,\delta^2(f)$, where the $\delta$ function
is normalized so that $\int |dz\,d\bar z|\delta^2(z)=1$. The
normalization ensures  that for any complex number $b$ and any
function $f(z)$, we have \eqn\torag{\int
dz\,\bar\delta(z-b)f(z)=f(b).}

Actually, though the  details will not be important for most
purposes, the wavefunction \murgo\ should be modified slightly to
have the right homogeneity in all variables.  Using the standard
transformation law of the delta function,  the object
$\bar\delta(\langle\lambda,\lambda_i\rangle)$ is homogeneous of
degree $-1$ in both $\lambda$ and $\lambda_i$; when we want to
make this explicit we write it as
$\bar\delta_{(-1,-1)}(\langle\lambda,\lambda_i\rangle)$.  On the
support of the delta function,  $\lambda$ is a nonzero multiple of
$\lambda_i$, so there is a well-defined and non-zero holomorphic
function $\lambda/\lambda_i$.  Hence we can define a more general
delta function
\eqn\rogo{\bar\delta_{(n-1,-n-1)}(\langle\lambda,\lambda_i\rangle)=
(\lambda/\lambda_i)^n
\bar\delta_{(-1,-1)}(\langle\lambda,\lambda_i\rangle).} The
wavefunction for a positive helicity gauge boson of momentum
$p_i=\lambda_i\tilde\lambda_i$ is actually
\eqn\pogo{\Psi^+_{i}(\lambda,\mu)=
\bar\delta_{(0,-2)}(\langle\lambda,\lambda_i\rangle)
\exp\left(i[\mu,\tilde\lambda_i](\lambda_i/\lambda)\right).} The
powers of $\lambda/\lambda_i$ have been included to ensure that
$\Psi^+$ is homogeneous of degree zero in $\lambda,\mu$.  It is
also homogeneous of degree $-2$ under
$(\lambda_i,\tilde\lambda_i)\to
(t\lambda_i,t^{-1}\tilde\lambda_i)$, and this ensures, as expected
(see section 2 of \witten\ for a review), that the scattering
amplitude for a gluon of positive helicity scales under that
transformation as $t^{-2}$. To write a twistor space wavefunction
for a gluon of the same momentum and negative helicity, we must
include the fermionic homogeneous coordinates $\psi^A$,
$A=1,\dots,4$ of $\Bbb{CP}^{3|4}$.  The wavefunction is
\eqn\logo{\Psi^-_{i}(\lambda,\mu,\psi)=\bar\delta_{(-4,2)}(\langle\lambda,\lambda_i\rangle)
\psi^1\psi^2\psi^3\psi^4\exp\left(i[\mu,\tilde\lambda_i](\lambda_i/\lambda)\right).}
The weights are chosen so that the wavefunction is homogeneous of
degree zero in overall scaling of $\lambda,\mu,$ and $\psi$.
Under $(\lambda_i,\tilde\lambda_i)\to
(t\lambda_i,t^{-1}\tilde\lambda_i)$, the wavefunction scales as
$t^2$, and that therefore is the scaling of the scattering
amplitude for a gluon of negative helicity. Again, this is the
standard result.

 To write a wavefunction for an external particle of
helicity $h=1-k/2$, we write  a similar formula with $k$ factors
of $\psi$. In addition, each external particle is also labeled by
an element $T_i$ of the Lie algebra of $U(N)$, which we have
omitted in writing the wavefunctions.

\subsec{ Curves In Twistor Space}

In twistor string theory, Yang-Mills scattering amplitudes are computed
by integration over the moduli space of holomorphic curves in
$\Bbb{CP}^{3|4}$.  Such a curve can be described as follows.

Begin with an abstract Riemann surface $C$ and a holomorphic line
bundle ${\cal L}$  (which must have enough nonzero holomorphic sections
or the following construction will be vacuous).  A holomorphic map
$\Phi:C \to\Bbb{CP}^{3|4}$ (which generically will be an
embedding) is described by picking sections $P_a(x)$, $Q_{\dot
a}(x)$ and $\chi^A(x)$ of ${\cal L}$ ($x$ denotes a point in $C$)
and setting \eqn\eqip{\eqalign{\lambda_a&=P_a(x) \cr
                   \mu_{\dot a}&=Q_{\dot a}(x)\cr
                         \psi^A&=\chi^A(x).\cr}}
 Geometrically, ${\cal L}=\Phi^*({\cal O}(1))$, where
${\cal O}(1)$ is the usual line bundle over $\Bbb{CP}^{3|4}$
(whose sections are functions homogeneous of degree one in the
homogeneous coordinates $\lambda,\mu,\psi$ of $\Bbb{CP}^{3|4}$),
and hence every holomorphic curve in $\Bbb{CP}^{3|4}$ arises by
this construction for some ${\cal L}$. Let ${\cal M}$ be the
moduli space of such curves; the moduli are the moduli of $C$ and
${\cal L}$ and the parameters that enter in picking the
polynomials $P_a$, $Q_{\dot a}$, and $\chi^A$.  The parameters in
the polynomials should be taken modulo an overall scaling, since a
common scaling of $P_a$, $Q_{\dot a}$, and $\chi^A$ does not
change $C$.

Each external gluon of momentum $p_i$ and wavefunction $\Psi_i$
couples to $C$ via an interaction $W_i=\int_C \Tr \Psi_i\wedge V$,
where $V$ is a vertex operator (in the worldsheet theory of the
$D1$-brane) that was described in \witten; the trace is taken in
the Lie algebra of $U(N)$. To compute the scattering amplitudes
for external gluons of momentum $p_i$ and  wavefunctions $\Psi_i$,
we must evaluate the integral \eqn\beqip{\int_{{\cal
M}}d\mu\,\left\langle \prod_i W_i\right\rangle} where $d\mu$ is a
suitable holomorphic measure, and the integral really is taken
over a suitable real cycle in ${\cal M}$.

There is no problem in integrating over the parameters in the
polynomials $P,Q$, and $\chi$.  Indeed, each of $P_a$, $Q_{\dot
a}$, and $\chi^A$ takes values in a common vector space
$U=H^0(C,{\cal L})$. Picking  an arbitrary basis $u_\sigma$,
$\sigma=1,\dots, r$ for this vector space, we expand
\eqn\anon{\eqalign{P_a & = \sum_\sigma p_{\sigma\,a}u_\sigma\cr
 Q_{\dot a} & = \sum_\sigma q_{\sigma\,\dot a}u_\sigma\cr
\chi^A & = \sum_\sigma \eta^A_{\sigma}\,u_\sigma.\cr}} A natural
measure for integrating over $P$, $Q$, and $\chi$ is then
\eqn\ononp{\Omega_0= \prod_{\sigma=1}^r\prod_{a,\,\dot a,\,A}
dp_{\sigma\,a}\,dq_{\sigma\,\dot a} \,d\eta^A_{\sigma}.} This
measure is independent of the choice of basis $u_\sigma$, since
the number of bosonic and fermionic variables is the same for each
$\sigma$. Since we really want to consider $P$, $Q$, and $\chi$ up
to a common scaling, we  want to
 integrate not over the space $\Bbb{C}^{4r|4r}$
of $P,Q$, and $\chi$, but over the corresponding projective space
$\Bbb{CP}^{4r-1|4r}$.  Being $\Bbb{C}^*$-invariant, the natural
measure $\Omega_0$ descends to an equally natural measure $\Omega$
on $\Bbb{CP}^{4r-1|4r}$.

If $C$ has genus zero, then $C$ and ${\cal L}$ have  no moduli, and hence
the measure just described can serve as a measure on ${\cal M}$.
For $C$ of positive genus, $C$ and ${\cal L}$ do have moduli.  It is not
at all clear what sort of measure should be used to integrate over those
moduli, so at the moment there is not a clear framework for higher genus
computations in twistor space.  For this reason, in section 3,
when we verify parity invariance of the scattering amplitudes in a precise
fashion, we consider only the contributions from curves of genus zero.
However, an important part of the calculation can be carried out
for arbitrary genus, as we will now see.

\subsec{Parity Symmetry And Duality}

Because of the factor
$\bar\delta(\langle\lambda,\lambda_i\rangle)$ in the wavefunction
$\Psi_i$ of the $i^{th}$ external gluon, this gluon is actually
attached to the curve $C$ at a point $x_i$ at which
\eqn\buvcu{\langle\lambda(x_i),\lambda_i\rangle = 0.}
Equivalently, since $\lambda_a(x)=P_a(x)$ along $C$, the condition is
\eqn\ucvu{\langle P(x_i),\lambda_i\rangle=0.}

Twistor space is constructed in a way that breaks the parity
symmetry between $\lambda_i$ and $\tilde\lambda_i$.
$\tilde\lambda_i$ enters the formalism quite differently.  As we saw in the formulas
of section 2.1 for the wavefunctions, the factor in $\Psi_i$ that
depends on $\tilde\lambda_i$ is $\exp\left(i[\mu,\tilde\lambda_i]
(\lambda_i/\lambda)\right).$ Since on $C$, $\mu_{\dot a}=Q_{\dot
a}(x)$, this factor actually becomes
$\exp\left(i[Q(x_i),\tilde\lambda_i](\lambda_i/ \lambda)\right).$
There is such a factor for each $i$, and these factors are the
only factors in the integrand of \beqip\ that depend on $Q$.  So
the integral over $Q$ reads \eqn\jippy{\int dQ\,\,\prod_i
\exp\left(i[Q(x_i), \tilde\lambda_i](\lambda_i/ \lambda)\right).}
Concretely, the integral over $Q$ is an integral $\int \prod_{\sigma\,\dot
a} dq_{\sigma\,\dot a}$ over all of the coefficients when
$Q$ is expanded in the basis $u_\sigma$. As in \rsv, we assume
that the integral should be taken as an integral over the real
axis. This being so, the integral over $Q$ simply gives a product
of delta functions, asserting that the amplitude receives its
contribution from curves $C$ and sets of points $x_i$ such that
\eqn\hippy{\sum_i[Q(x_i),\tilde\lambda_i](\lambda_i/\lambda(x_i))=0}
for every $Q\in H^0(C,{\cal L})$.

Now we can see why parity symmetry is a problem: the curves $C$ and
sets of points $x_i$ that contribute to the scattering amplitude
are constrained by the set of equations \ucvu\ and \hippy\ in
which $\lambda_i$ and $\tilde\lambda_i$ enter quite
asymmetrically.  Our goal is to express these conditions in a
symmetrical fashion.

To do this, we let $K$ denote the canonical line bundle of $C$,
and for each point $x_i\in C$, we let $\CO(x_i)$ denote the line
bundle whose holomorphic sections are holomorphic functions on $C$
that are allowed to have a simple pole at $x_i$.   The line bundle
$K(x_1,\dots, x_n)=K\otimes\left(\otimes_{i=1}^n\CO(x_i)\right)$
(which we also abbreviate as $K(x_i)$)
has for its holomorphic sections the holomorphic differentials on
$C$ that may have simple poles at the $x_i$ (and no other
singularities). Such a differential $\omega$ has, at each point $x_i$
where there may be a pole, a {\it residue}, a complex number
$c_i={\rm Res}_{x_i}\omega$.

Suppose instead that $\omega$ is a holomorphic section of
$K(x_i)\otimes {\cal L}^{-1}=K\otimes {\cal
L}^{-1}\otimes(\otimes_i\CO(x_i)).$ Thus, $\omega$ is now a holomorphic
differential with values in ${\cal L}^{-1}$ that may have poles at
the $x_i$. We can still define the residues $c_i={\rm
Res}_{x_i}\omega$, but now the $c_i$, instead of being complex
numbers, take values in the vector spaces ${\cal L}^{-1}_{x_i}$,
the fibers of ${\cal L}^{-1}$ at $x_i$.\foot{ Locally, near $x_i$, we
can trivialize ${\cal L}^{-1}$; once this is done, $\omega$ is an
ordinary differential form with a possible pole at $x_i$, and its
residue is a complex number. This number depends on how ${\cal
L}^{-1}$ was trivialized; the intrinsic description is that $c_i$
is a vector in the one-dimensional vector space ${\cal
L}^{-1}_{x_i}$.}

Suppose that, for $\dot a=1,2$,
 we can find   $\tilde P_{\dot a}\in H^0(C,K(x_i)\otimes {\cal L}^{-1})$
 such that, for all $i$, \eqn\inon{\tilde\lambda_{i\,\dot
a}(\lambda_i/\lambda(x_i)) = {\rm Res}_{x_i}\tilde P_{\dot a}.}
Notice that the left hand side of \inon\ takes values in ${\cal
L}^{-1}_{x_i}$, because of the appearance of $\lambda(x_i)$ in the
denominator. Thus, the left and right hand sides of \inon\ take
values in the same vector space, and the equation makes sense.

If so, let $\omega=Q^{\dot a}\tilde P_{\dot a}$, for any $Q^{\dot
a}\in H^0(C,{\cal L})$. As $Q^{\dot a},$ $\dot a=1,2$, is a
holomorphic section of ${\cal L}$, $\omega$ is a section of
$K(x_i)$ and thus can be interpreted as an ordinary holomorphic
differential on $C$ with possible simple poles at the $x_i$.  Its
residues are therefore ordinary complex numbers, which simply
equal $Q^{\dot a}\tilde\lambda_{i\,\dot
a}(\lambda_i/\lambda(x_i))$ (indeed, the residue of
$\omega=Q\tilde P$ at $x_i$ is the value there of $Q$, which has
no pole at $x_i$, times the residue of $\tilde P$). The usual
residue theorem asserts that the sum of these residues vanishes:
\eqn\nondon{\sum_iQ^{\dot a}(x_i) \tilde\lambda_{i,\dot
a}(x_i)(\lambda_i/\lambda(x_i)) =0.} But this is precisely the
desired condition \hippy!

Below, we will show that, conversely, \hippy\ is satisfied only if
there exists a differential $\tilde P_{\dot a}$ obeying \inon.
Assuming this for a moment, we can now restate the basic
conditions \ucvu\ and \hippy\ to treat $\lambda_i$ and
$\tilde\lambda_i$ symmetrically. We set  $\tilde{\cal L}=K(x_i)\otimes
{\cal L}^{-1}$. Thus \eqn\sonno{{\cal L}\otimes \tilde {\cal L}=K(x_i).}
The right hand side of the last formula depends only on the choice
of the curve $C$ and the points $x_i\in C$; there is no asymmetry
here between $\lambda$ and $\tilde\lambda$. The left hand side is
symmetrical in ${\cal L}$ and $\tilde{\cal L}$; when we exchange
$\lambda$ and $\tilde\lambda$, we will also exchange ${\cal L}$
and $\tilde{\cal L}$.

The basic equations obtained so far relating $\lambda$,
$\lambda_i$, and $\tilde\lambda_i$ to $P_a$ and $\tilde P_{\dot
a}$ are as follows: \eqn\ughu{
\eqalign{\lambda_a(x_i)& = P_a(x_i)
\cr
                   \langle\lambda(x_i),\lambda_i\rangle& = 0\cr
                    \tilde\lambda_{i\,\dot a}(\lambda_i/\lambda(x_i))&
={\rm Res}_{x_i}\tilde P_{\dot a}.\cr}} The second of these
equations asserts that $\lambda_i$ is a multiple of
$\lambda(x_i)$, so $(\lambda_i/\lambda)$ is a well-defined complex
number, as is assumed in writing the third equation. Also, it
follows from the first two equations that
\eqn\bughu{\lambda_{i\,a}=w_iP_a(x_i)} for some $w_i$ (which takes
values in ${\cal L}_{x_i}^{-1}$). The third equation in \ughu\
similarly implies that \eqn\tughu{\tilde\lambda_{i\,\dot a}=\tilde
w_i\,\tilde P_{\dot a}(x_i)} for some $\tilde w_i$.

Eqn. \tughu\ may require some elucidation. There are two ways to
think about $\tilde P_{\dot a}$. If it is viewed as a section of
$K\otimes {\cal L}^{-1}$ that has a pole at $x_i$, then it has a
residue $r_{i\,\dot a}$ at $x_i$ which takes values in $ {\cal
L}^{-1}$. This is the point of view we used so far; it enabled us
to invoke the residue theorem to derive \nondon. A different point
of view is more helpful for understanding the symmetry between
$\lambda$ and $\tilde\lambda$.   If we view $\tilde P_{\dot a}$
simply as a section of a line bundle $K(x_i)\otimes {\cal
L}^{-1}$, then its ``value'' at $x_i$ is an element which we call
$\tilde P_{\dot a}(x_i)$ of the fiber of this line bundle at
$x_i$; this fiber is $(K(x_i)\otimes {\cal
L}^{-1})_{x_i}=K(x_i)_{x_i}\otimes {\cal L}^{-1}_{x_i}$.  The
relation between the two points of view comes from the fact that,
with $K(x_i)$ understood as the line bundle whose sections are
holomorphic differentials with possible simple poles at the $x_i$,
the fiber $K(x_i)_{x_i}$ is naturally trivial.\foot{Our notation
here is really too compressed.  We recall that $K(x_i)$ is an
abbreviation for $K(x_1,\dots,x_n)=K\otimes(\otimes_i\CO(x_i))$.
By $K(x_i)_{x_i}$ we mean $K(x_1,\dots,x_n)_{x_i}$, that is, the
fiber of $K(x_1,\dots,x_n)$ at $x_i$. A partial justification of
our overly compressed notation is that for analyzing the fiber of
$K(x_1,\dots,x_n)$ at $x_i$, the existence and location of the
$x_j$ with $j\not= i$ are irrelevant.  Thus the fiber of
$K(x_1,\dots,x_n)$ at $x_i$ is naturally isomorphic to the fiber
of $K\otimes \CO(x_i)$ at $x_i$; the latter is perhaps a better
candidate for being called $K(x_i)_{x_i}$.} The trivialization is
made by the residue map; if $\omega$ is a holomorphic differential
with a simple pole at $x_i$, then the coefficient of the pole,
which is the value of $\omega$ at $x_i$, is in a natural way a
complex number, namely ${\rm Res}_{x_i}\omega$.  Going back to our
problem, what in one point of view is called ${\rm
Res}_{x_i}\tilde P_{\dot a}$, an element of ${\cal L}^{-1}_{x_i}$,
is in the other point of view simply called $\tilde P_{\dot
a}(x_i)$, an element of $(K(x_i)\otimes {\cal L}^{-1})_{x_i}$. The
two points of view are compatible because the latter space is
isomorphic by the residue map to $ {\cal L}^{-1}_{x_i}$.

This hopefully makes it clear that \tughu\ is equivalent to the
third equation in \ughu, with $\tilde w_i=(\lambda(x_i)/
\lambda_i)$. Since $\lambda_{i\,a}=w_iP_a=w_i\lambda_a(x_i)$, we
similarly have $w_i=(\lambda_i/\lambda(x_i))$. Thus $w_i\tilde
w_i=1$.  We cannot expect to find any further constraints on $w_i$
and $\tilde w_i$, because we are free to rescale $\lambda_i\to
t_i\lambda_i$, $\tilde\lambda_i\to t_i^{-1}\tilde\lambda_i$, for
any $t_i\in \Bbb{C}^*$.  Clearly all the conditions that we have
described are completely symmetric in $\lambda$ and
$\tilde\lambda$, establishing the parity invariance of the
formalism.

For an alternative view of things,
 we  combine the three
equations in \ughu\ to write \eqn\bughu{\lambda_{i\,
a}\tilde\lambda_{i\,\dot a}=P_a(x_i){\rm Res}_{x_i}\tilde P_{\dot
a}.} Now let \eqn\ononn{\omega_{a\dot a}=P_a\tilde P_{\dot a}.}
Each component of $\omega_{a\dot a}$, $a,\dot a=1,2$, is a section
of $K(x_i)$; that is, it is an ordinary holomorphic differential
with possible simple poles at the $x_i$. The factorization \ononn\
implies that \eqn\tononn{\omega_{a\dot a}\omega^{a\dot a}=0.} And
finally, we have \eqn\nughu{\lambda_{i\,a}\tilde\lambda_{i\,\dot
a}={\rm Res}_{x_i}\omega_{a\dot a}.} These conditions are actually
closely related to the saddle point equation \ref\grossmende{D. J.
Gross and P. F. Mende, ``String Theory Beyond The Planck Scale,''
Nucl. Phys. {\bf B303} (1988) 407.} that describes high energy,
fixed angle scattering in string perturbation theory.  The
relation will be further explored elsewhere.

Clearly, equations \tononn\ and \nughu\ are symmetrical in
$\lambda$ and $\tilde\lambda$.  Moreover, all the structure
described previously can be deduced from those equations. For
example, \tononn\ says that at any given $x\in C$, $\omega_{a\dot
a}$ is a null vector and so has a factorization $\omega_{a\dot
a}=P_a\tilde P_{\dot a}$, unique up to $P\to tP$, $\tilde P\to
t^{-1}\tilde P$. To make this factorization globally, we must
interpret $P$ and $\tilde P$ as sections of suitable line bundles,
reasoning as follows. We define a complex quadric
$W\subset\Bbb{CP}^3$ that has homogeneous complex coordinates
$Z_{a\dot a}$, $a,\dot a=1,2$ obeying $Z_{a\dot a}Z^{a\dot a}=0$.
As long as the $\omega_{a\dot a}$ have no common zeroes, which is
true generically, we can define a map $\Phi:C\to W$ by setting
$Z_{a\dot a}=\omega_{a\dot a}$; this maps $C$ to $W$ (and not just to $\Bbb{CP}^3$)
 since
$\omega_{a\dot a}\omega^{a\dot a}=0$. Since the $Z_{a\dot a}$ are
homogeneous coordinates, $\Phi$ is well-defined even at poles of
$\omega_{a\dot a}$ (if $\omega_{a\dot a}$ has a pole at $x=x_i$,
one can define $\Phi$ near $x_i$ by $Z_{a\dot
a}=(x-x_i)\omega_{a\dot a}$).
 The quadric $W$ is isomorphic as a complex manifold to
$\Bbb{CP}^1\times \Bbb{CP}^1$; there are two natural line bundles
$\CO(1)$ and $\CO(1)'$ over it, and the $Z_{a\dot a}$ are elements
of $H^0(W,\CO(1)\otimes\CO(1)')$.  Moreover, there are sections
$\lambda^W_a\in H^0(W,\CO(1))$ and $\tilde\lambda^W_{\dot a}\in
H^0(W,\CO(1)')$ with $Z_{a\dot a}=\lambda^W_a\tilde\lambda^W_{\dot
a}$.\foot{The quickest way to prove these assertions is to start
with $\Bbb{CP}^1\times \Bbb{CP}^1$ and let $\lambda_a^W$ and
$\tilde\lambda_{\dot a}^W$ denote the homogeneous coordinates of,
respectively, the first and second factor (so they are the
holomorphic sections of $\CO(1)$ and $\CO(1)'$, respectively).
Then  simply define a map from $\Bbb{CP}^1\times \Bbb{CP}^1$ to
$\Bbb{CP}^3$ by $Z_{a\dot a}=\lambda_a^W\tilde\lambda^W_{\dot a}$.
Clearly this maps $\Bbb{CP}^1\times \Bbb{CP}^1$ to the quadric
$Z_{a\dot a}Z^{a\dot a}=0$; it is easily seen to be an
isomorphism.} Finally, to recover the formalism that we found
above, we set ${\cal L}=\Phi^*(\CO(1))$, $\tilde{\cal
L}=\Phi^*(\CO(1)')$, $P_a=\Phi^*(\lambda^W_a)$, and $\tilde
P_{\dot a}=\Phi^*(\tilde\lambda^W_{\dot a})$.

\bigskip\noindent{\it The Converse Statement}

We still must prove the converse statement that \hippy\ holds only if
$\tilde\lambda_{i\,\dot a}$ can be expressed as in \inon\
in terms of a suitable differential
$\tilde P_{\dot a}$.

We consider the exact sequence of sheaves \eqn\unvup{0\to K\otimes
{\cal L}^{-1}\underarrow{i} K(x_i)\otimes {\cal L}^{-1}
\underarrow{r} \oplus_i  {\cal L}^{-1}_{x_i}\to 0.} By $K\otimes
{\cal L}^{-1}$ we mean the sheaf of sections of the line bundle
$K\otimes {\cal L}^{-1}$, and similarly for $K(x_i)\otimes{\cal
L}^{-1} $. Also, $\oplus_i  {\cal L}^{-1}_{x_i}$ is the sheaf
whose sections are families $\{\alpha_i\}$ with each $\alpha_i$ a
vector in ${\cal L}^{-1}_{x_i}$.  The map $r$ is the ``residue''
map which maps a holomorphic section $\phi$ of $K(x_i)\otimes{\cal
L}^{-1}$, which we recall is a holomorphic section of $K\otimes
{\cal L}^{-1}$ with possible poles at the $x_i$, to the family
$\{\alpha_i\}$ where $\alpha_i={\rm Res}_{x_i}(\phi)$. The map $i$
is the ``inclusion'' of sheaves, which maps a holomorphic section
$\rho$ of $K\otimes {\cal L}^{-1}$ to the ``same'' section of
$K(x_i)\otimes {\cal L}^{-1}$. $i(\rho)$ does not have poles at
$x_i$, and so has vanishing residues; hence $ri(\rho)=0$.
Conversely, if a section $\phi$ of $K(x_i)\otimes{\cal L}^{-1}$ is
annihilated by $r$, that is, it has no poles at $x_i$, then it can
be regarded as a section of $K\otimes {\cal L}^{-1}$, and hence is
of the form $i(\rho)$ for some $\rho$. These assertions are part
of the statement that the sequence \unvup\ is exact.  The
remainder of the statement of exactness is the assertion that $i$
is injective, which is obvious, and that $r$ is surjective, which
expresses the fact that locally the residues of a differential can
be specified arbitrarily.

The short exact sequence of sheaves \unvup\ leads to a long exact
cohomology sequence which reads in part \eqn\tunvup{\dots
H^0(C,K(x_i)\otimes {\cal L}^{-1})\underarrow{r} \oplus_i{\cal
L}^{-1}_{x_i} \underarrow{\delta} H^1(C,K\otimes {\cal
L}^{-1})\dots.} In our problem, we have a family $\{\alpha_{i}\}$,
with $\alpha_{i}=\tilde\lambda_{i\, \dot a}(\lambda_i/\lambda)$
(in this discussion we  regard $\dot a$ as a fixed number, 1 or
2), and we want to know if there is a global differential $\tilde
P$ such that $\alpha_{i}={\rm Res}_{x_i}\tilde P_{i}$. The
exactness of the sequence \tunvup\ asserts that $\tilde P$ exists
if and only if $\delta(\{\alpha_{i}\})=0$. The definition of the
map $\delta$ is that $\delta(\{\alpha_i\})$ is an element of
$H^1(C,K\otimes{\cal L}^{-1})$ that can be represented by the
$K\otimes {\cal L}^{-1}$-valued $(0,1)$-form
$\zeta=\sum_i\alpha_idx\bar\delta(x-x_i)$, where $x$ is an
arbitrary local holomorphic parameter near $x=x_i$ (and as
explained in section 2.1, $\bar\delta(x-x_i)=id\bar
x\delta^2(x-x_i)).$ One can verify, using the transformation of
the delta function under a change of coordinates, that $\zeta$ is
independent of the choices of local coordinates.

$\zeta$ is nonzero as a differential form, but we need to know if
it is nonzero as an element of $H^1(C,K\otimes {\cal L}^{-1})$.
For this, we can use Serre duality, where asserts that
$H^1(C,K\otimes {\cal L}^{-1})$ is the dual space to $H^0(C,{\cal
L})$, and more precisely that an element $\zeta\in
H^1(C,K\otimes{\cal L}^{-1})$ vanishes if and only if $\int_C
\zeta Q=0$ for every $Q\in H^0(C,{\cal L})$. In our case, because
of the delta functions in the definition of $\zeta$, the integral
is trivially done: $\int_C\zeta Q=\sum_i\alpha_i Q(x_i)$. Putting
it all together, we have established what we wanted to know: the
family $\{\alpha_i\}$ can be written as ${\rm Res}_{x_i}\tilde P$,
for some global differential $\tilde P$, if and only if
$\sum_i\alpha_i Q(x_i)=0$ for every $Q\in H^0(C,{\cal L})$.

This completes the demonstration that the twistor representation of the scattering
amplitude can be expressed by integration over a left-right symmetric set of
 parameters --  $C$, the $x_i$, ${\cal L}$ and $\tilde{\cal L}$, and
$P_a$ and $\tilde P_{\dot a}$.

\newsec{Explicit Evaluation For Genus Zero}

In genus zero, we can make this much more explicit.  Moreover, because the
integration measure is known in genus zero, we can give a complete proof of
parity invariance of the tree level scattering amplitudes, by showing that after
the manipulation explained in section 2, the
integration measure as well as the integration region becomes left-right symmetric.

To compute the scattering amplitude, we must integrate over the
choice of curve $C$ and line bundle ${\cal L}$, over the points
$x_i\in C$ at which the external gluons are attached, and over the
polynomials $P_a$, $Q_{\dot a}$, and $\chi^A$.  In genus zero, $C$
and ${\cal L}$ have no moduli.  The integration measure for the
polynomials was explained in section 2. Finally, the integration
measure over the $x_i$ comes from the path integral of free
fermions on the worldvolume of the $D1$-brane, as explained in
\witten.  We describe $C\cong \Bbb{CP}^1$ by homogeneous
coordinates $u^\alpha$, $\alpha=1,2$, and write $u_i^\alpha$ for
the homogeneous coordinates of $x_i$. For a single trace
subamplitude with external gluons attached in a definite cyclic
order (which we take to be simply $123\dots n$), the measure that
comes from the worldvolume path integral is \eqn\nonno{\prod_i
\int \langle u_i,du_i\rangle {1\over \prod_k\langle
u_k,u_{k+1}\rangle}.} The wave functions all contain a factor
$\bar\delta(\langle P(u_i),\lambda_i\rangle)$, which is supported
for $u_i$ such that $P(u_i)$ is a multiple of $\lambda_i$. Which
multiple it is does not matter, since \nonno\ is homogeneous in
$u_i$, for each $i$.  It is convenient to take the multiple to be
1, which we do by using the fact that\foot{If the argument of the
delta function on the left vanishes at $u_i=\alpha_i$, for some
$\alpha_i$, then by homogeneity it vanishes at $u_i=w\alpha_i$ for
any $w$.  However, on the left, $u_i$ is a homogeneous variable
and we can just set $w=1$.  Instead, on the right, as $P$ is
homogeneous of degree $q-1$, there are $q-1$ values of $w$ for
which the argument of the delta function vanishes; however, each
contributes to the integral with a factor of $1/(q-1)$ that comes
from the fact that if $f(x)$ vanishes at $x=x_0$, the contribution
of this zero to $\int dx\, \delta(f(x))$ is $1/|f'(x_0)|$.}
\eqn\jonno{\eqalign{\prod_i &\int \langle u_i,du_i\rangle {1\over
\prod_k\langle u_k,u_{k+1}\rangle} \prod_m\bar\delta(\langle
P(u_m),\lambda_m\rangle)\cr &= \prod_i \int d^2u_i {1\over
\prod_k\langle u_k,u_{k+1}\rangle}
\prod_m\bar\delta^2(P_a(u_m)-\lambda_{m\,a}).\cr}} One advantage
of taking the multiple to be 1 is that we can drop all factors of
$(\lambda_i/\lambda)$ in the wavefunctions of section 2.  Upon
doing so, the wavefunctions look much more appealing.

We suppose that $p$ of the external gluons have positive helicity
and $q$ have negative helicity, with $p+q=n$. In this case, according to the rules
in \witten,  the
line bundle ${\cal L}$ is ${\cal L}=\CO(q-1)$.  The gluons with
right-handed or positive helicity form a set $R$, and the gluons
with left-handed or negative helicity form a set $L$. The
polynomials $P_a$, $Q_{\dot a}$, and $\chi^A$ are of degree $q-1$
in the $u^\alpha$. The space $U$ of degree $q-1$ polynomials
$f(u^\alpha)$ is $q$-dimensional. Taking advantage of the fact
that there are precisely $q$ points in $L$, we pick a basis for
$U$ consisting of basis elements $f_i$, $i\in L$, such that for
all $i,j\in L$, $f_i(u_j)=\delta_{ij}$.
 ($f_i$ is only defined for $i\in L$ and $f_i(u_j)$ is only constrained
to equal $\delta_{ij}$ if in addition  $j\in L$.) Explicitly (though we will never
use this formula), this
is accomplished by taking \eqn\imop{f_i(u)=\prod_{j\in L,\,j\not=
i}{\langle u,u_j\rangle\over \langle u_i,u_j\rangle}.}

Our general recipe for the integration measure says that we should
expand (for example) $P_a=\sum_{i\in L}p_{i\,a}f_i$, whereupon the
integration measure is \eqn\bonon{\prod_a\,dP_a=
\prod_{i,a}dp_{i\,a}.} With our choice of basis,
$p_{i\,a}=P_a(u_i)$ for $i\in L$, and we can alternatively write
the integration measure as \eqn\onnonn{\prod_{i\in L,a}dP_a(u_i).}
The integration measures for the other polynomials $Q$ and $\chi$
is precisely analogous and can be written as in \bonon\ or
\onnonn.

The combined integration measure for $P,Q$, and $\chi$ is
independent of the choice of basis, but picking a basis enables us
to integrate over $P,$ $Q$, or $\chi$ separately. This has several
benefits.  The first is that we can trivially integrate over
$\chi$ and eliminate the fermions from the discussion.  As we
explained in section 2, for every $i\in L$, the external gluon
wavefunction contains a factor $\prod_{A=1}^4\psi^A$.  For the
$i^{th}$ external gluon, $\psi^A=\chi^A(u_i)$, so the dependence
of the integrand on the $\chi^A$ is \eqn\onno{\prod_{i\in
L}\prod_{A=1}^4\chi^A(u_i).} On the other hand, with our choice of
basis, the measure for integrating over $\chi$ is
$\prod_{i,A}d\chi^A(u_i)$.  The definition of fermion integration
gives immediately \eqn\bonno{\int
\prod_{i,A}d\chi^A(u_i)\,\prod_{j\in
L}\prod_{B=1}^4\chi^B(u_j)=1,} so with our choice of basis, the
fermion integral simply gives a factor of 1.

As described in section 2.2, the scattering amplitude also
contains a factor \eqn\plooky{\int dQ_{\dot
a}\exp\left(i\sum_j[Q(u_j),\tilde\lambda_j]\right).} (As noted following eqn.
\jonno, we can drop factors of $\lambda/\lambda_j$.  This is done
below without comment.) We saw in section 2.3 that the integral is
supported on the locus on which $u_j$ and $\tilde\lambda_j$ are
such that \eqn\jind{\tilde\lambda_{j\,\dot a}={\rm
Res}_{u_j}\tilde P_{\dot a}} for some $\tilde P_{\dot a}\in
H^0(C,K(u_i)\otimes{\cal L}^{-1})$. We can write explicitly
\eqn\torry{\tilde P_{\dot a}=\langle u,du\rangle{T_{\dot
a}(u)\over \prod_i\langle u_i,u\rangle},} where $T_{\dot a}(u)$ is a polynomial
homogeneous of degree $p-1$, or in other words is a section of
$\tilde{\cal L}=\CO(p-1)$. This ensures that $\tilde P_{\dot a}$
is a differential homogeneous of degree $1-q$ with possible simple
poles at the $u_i$ or in other words a section of $K(u_i)\otimes
{\cal L}^{-1}$.  Calculating the residues, we can rewrite \jind\
as \eqn\bind{\tilde\lambda_{j,\dot a}={T_{\dot a}(u_j)\over
\prod_{k\not= j}\langle u_k,u_j\rangle}.}

The scattering amplitude will be expressed as an integral over
$T_{\dot a}$.  We pick a basis $\tilde f_\tau$ of $\tilde
U=H^0(C,\tilde{\cal L})$ and expand $T_{\dot a}$ in this basis:
$T_{\dot a}=\sum_\tau t_{\tau\,\dot a}\tilde f_\tau $.  The
integration measure over $T_{\dot a}$ is then taken to be
\eqn\yuro{\prod_{\tau,\,\dot a}dt_{\tau\,\dot a}.} Aiming for
left-right symmetry, we define the basis of $\tilde U$ in a
``dual'' fashion to the basis of $U$ that was chosen earlier. As
$\tilde U$ is $p$-dimensional, we introduce one basis element
$\tilde f_i$ for each $i\in R$, normalized so that $\tilde
f_i(u_j)=\delta_{ij}$ for $j\in R$. Explicitly (though again we
will not use this formula), \eqn\utry{\tilde f_i(u)=\prod_{j\in
R,\,j\not= i}{\langle u,u_j\rangle\over \langle u_i,u_j\rangle}.}
The integration measure then becomes \eqn\newyuro{\prod_{i\in
R,\,\dot a} dT_{\dot a}(u_i),} in perfect parallel with the
integration measure for $P$ and $Q$ as described earlier (see
\onnonn).

We now expect that \eqn\ufgo{\int dQ_{\dot
a}\exp\left(i\sum_j[Q(u_j),\tilde\lambda_j]\right)=g(u_1,\dots,u_n)\int
dT_{\dot a}\prod_j\delta\left(\tilde\lambda_{j\,\dot a}-{T_{\dot
a}(u_j)\over \prod_{k\not= j}\langle u_k,u_j\rangle}\right)} for
some function $g(u_j)$.  This expresses the fact that, as we know
from section 2.3, the integral on the left has delta function
support on the locus on which \bind\ is obeyed for some $T_{\dot
a}$.

To determine $g(u_j)$, we act on both the left and right hand side
with $\prod_{j\in L}\int d^2\tilde\lambda_{j,\dot a}$.  The
integration contour is taken to be the real axis.  On the left
hand side of \ufgo, the part of the exponent that involves
$\tilde\lambda_j$ for  $j\in L$ is just $\sum_{j\in
L}[q_j,\tilde\lambda_j]$. (The $q_j$ were defined so that
$Q(u_j)=q_j$ for $j\in L$ and the measure in integrating over $Q$
is just $\prod_{j\in L,\dot a}dq_{j\,\dot a}$.)   This being so,
we get \eqn\ifo{\prod_{j\in L}\int d^2\tilde\lambda_{j,\dot a}
\int dQ_{\dot
a}\exp\left(i\sum_j[Q(u_j),\tilde\lambda_j]\right)=(2\pi)^{2p},}
where each integral over $\tilde\lambda_{j\,\dot a}$ gives a
factor $2\pi\delta(q_{j\,\dot a})$, and the integral over $Q$ is
done with these delta functions.  To apply $\prod_{j\in L}\int
d^2\tilde\lambda_{j,\dot a}$ to the right hand side of \ufgo, we
simply note that \eqn\dufgo{\prod_{j\in L}\int
d^2\tilde\lambda_{j,\dot a}\int dT_{\dot
a}\prod_m\delta\left(\tilde\lambda_{m\,\dot a}-{T_{\dot
a}(u_m)\over \prod_{k\not= m}\langle u_k,u_m\rangle}\right) =\int
dT_{\dot a}\prod_{m\in R}\delta\left(\tilde\lambda_{m\,\dot
a}-{T_{\dot a}(u_m)\over \prod_{k\not= m}\langle
u_k,u_m\rangle}\right),} where all we have done is to evaluate the
integrals over the $\tilde\lambda_j$ for $j\in L$ using the delta
functions for $m\in L$.  The delta functions that remain are
therefore the ones for $m\in R$, and
 these, with our choice of basis, give
simply \eqn\loopy{\prod_{k\in R}\int dt_{k\,\dot a}\prod_{m\in
R}\delta\left(\tilde\lambda_{m\,\dot a}-{t_{m\,\dot a}\over
\prod_{j\not= m} \langle u_j,u_m\rangle}\right)
 =  \prod_{k\in R,\,m\not= k}\langle u_m,u_k\rangle^2.}
On the right hand side, all factors are squared simply because
each $t_k$ has two components $t_{k\,\dot a}$, $\dot a=1,2$; $k$
is restricted to $R$ but $m$ ranges over both $L$ and $R$.

Comparing these formulas, we see that
\eqn\udur{g(u_1,\dots,u_n)=(2\pi)^{2p} { \prod_{k\in R,\,m\not=
k}\langle u_m,u_k\rangle^{-2}}.} Now, we assume that the
underlying scattering amplitude is defined as \eqn\urono{{1\over
(2\pi)^{2p}}\int dP\,dQ\,d\chi\, \left\langle \prod_i
W_i\right\rangle} where as in \beqip, $\left\langle \prod_i
W_i\right\rangle$ is the correlation function of vertex operators
in the worldvolume theory of the $D$-instanton, and the integral
over $P,$ $Q$, and $\chi$ is the integral over the moduli space
${\cal M}$ of $D$-instantons, as described in \witten\ and in
section 2.  The only novelty we are adding here relative to what
was explained in \witten\  is that a factor of $(2\pi)^{-2p}$ must
be included in the definition in order to ensure parity symmetry.
(This can be interpreted as a normalization factor in the
wavefunctions or the gauge coupling constant.)  We have seen that
with our choice of basis, the integral over $\chi$ gives 1, and
the integral over $Q$ can be replaced by an integral over $T$ with
the extra factor $g(u_i)$. It also was shown in \witten\ that the
evaluation of the correlation function gives $\prod_j\int \langle
u_j,du_j\rangle \prod_m\langle u_m,u_{m+1}\rangle^{-1}$, and in
eqn. \jonno\ we explained how to convert $\int \langle
u,du\rangle$ to $\int d^2 u$. (However, we will here take all
integration variables, including $u$, to be real and replace
$\bar\delta$ functions by ordinary delta functions.)  So all told
upon replacing the integral over $Q$ by an integral over $T$ via
the above-described recipe, the formula for the scattering
amplitude becomes \eqn\proofgo{\eqalign{A= \int dP_a\,dT_{\dot
a}\int\prod_j d^2u_j &{1\over \prod_h\langle u_h,u_{h+1}\rangle}
{1\over \prod_{k\in R,\,m\not= k}\langle u_m,u_k\rangle^2}\cr &
\prod_{i}\delta^2\left(\lambda_{i\,a}-P_a(u_i)\right)
\delta^2\left(\tilde\lambda_{i\,\dot a}-{T_{\dot a}(u_i)\over
\prod_{l\not= i}\langle u_l,u_i\rangle}\right).\cr}}

We have not quite achieved manifest parity-invariance, but as will
soon be clear, this can now be obtained by an elementary change of
variables. Let $\phi_i=1$ for $i\in L$ and 0 for $i\in R$, and let
$\tilde\phi_i=1-\phi_i$.  Then as we will see, \proofgo\ is
equivalent to the manifestly symmetric expression
\eqn\roofgo{\eqalign{A= \int dP_a\,dT_{\dot a}&\int\prod_j d^2u_j
{1\over \prod_m\langle u_m,u_{m+1}\rangle} {1\over \prod_{l\not=
k}\langle u_l,u_k\rangle^2}\cr
 &
\prod_{i}\delta^2\left(\lambda_{i\,a}-{P_a(u_i)\over \prod_{s\not=
i}\langle u_s, u_i\rangle^{\phi_i}}\right)
\delta^2\left(\tilde\lambda_{i\,\dot a} -{T_{\dot a}(u_i)\over
\prod_{t\not= i}\langle
u_t,u_i\rangle^{\tilde\phi_i}}\right).\cr}} There are a total of
$4n$ integration variables and $4n$ delta functions, so the
integral really reduces to a sum over contributions of delta
functions.

We make the change of variables \eqn\incoming{\eqalign{u_i & \to
u_i\prod_{k\not= i}\langle u_k,u_i\rangle^{-1/(q-1)},~~i\in L \cr
u_i&\to u_i,~~i\in R\cr P& \to P\cr
 T&\to {T\over \prod_{j\in L,\,k\not=j}\langle
 u_k,u_j\rangle^{1/(q-1)}}.\cr}}
We find that \eqn\iko{\eqalign{P(u_i)&\to {P(u_i)\over
\prod_{k\not=i}\langle u_k,u_i\rangle},~~i\in L\cr P(u_i)&\to
P(u_i),~~i\in R.}} These formulas show that the delta functions
containing $P$ in \proofgo\ transform into the delta functions
containing $P$ in \roofgo.  Also, since the measure for the $P$
integration is $dP_a=\prod_{i\in L}\prod_ad^2P_a(u_i)$, the first
formula in \incoming\ shows that this measure transforms as
\eqn\onsonson{\prod_a dP_a\to{\prod_a dP_a\over \prod_{i\in
L,\,j\not=i}\langle u_j,u_i\rangle^2}.} (The denominator is
squared because we must transform $P_a$ for $a=1,2$.)  We also
have \eqn\omorf{\eqalign{T(u_i)\over\prod_{j\not= i}\langle
u_j,u_i\rangle & \to T(u_i),~~i\in L\cr T(u_i)\over\prod_{j\not=
i}\langle u_j,u_i\rangle& \to {T(u_i)\over\prod_{j\not= i}\langle
u_j,u_i\rangle},~~i\in R.\cr}} These formulas imply that the delta
functions containing $T$ in \proofgo\ transform into the delta
functions containing $T$ in \roofgo.  Furthermore, since the
measure for integrating over $T$ is $dT_{\dot a}=\prod_{i\in
R}\prod_{\dot a} dT_{\dot a}(u_i)$, the second formula shows that
\eqn\uncuvo{{\prod_{\dot a}dT_{\dot a}\over \prod_{i\in R,\,j\not=
i}\langle u_j,u_i\rangle^2}} is invariant under the rescaling.
Finally, \eqn\ucju{{\prod_i d^2u_i\over \prod_j\langle
u_j,u_{j+1}\rangle}} is invariant under any scaling of the $u_i$,
and in particular under the transformation in \incoming.  If we
combine these assertions, we find that the given change of
variables does indeed transform \proofgo\ into \roofgo.

\bigskip
This work was supported in part by NSF Grant PHY-0070928.  I thank
N. Berkovits, F. Cachazo and P. Svrcek for discussions. \listrefs
\end